\documentclass[]{spie}  

\usepackage{amsmath,amsfonts,amssymb}
\usepackage{graphicx}
\usepackage[colorlinks=true, allcolors=blue]{hyperref}

\title{Interactive Layout Drawing Interface with Shadow Guidance}

\author[a]{Jiahao Weng}
\author[a]{Haoran Xie}
\affil[a]{Japan Advanced Institute of Science and Technology, Ishikawa, Japan}

\authorinfo{Further author information: (Send correspondence to H. Xie)\\J.Weng: E-mail: s2110022@jaist.ac.jp\\  H. Xie: E-mail: xie@jaist.ac.jp, Telephone: +81-0761-51-1753}

\begin{document} 
\maketitle

\begin{abstract}
It is difficult to design a visually appealing layout for common users, which takes time even for professional designers. In this paper, we present an interactive layout design system with shadow guidance and layout retrieval to help users obtain satisfactory design results. This study focuses in particular on the design of academic presentation slides. The user may refer to the shadow guidance as a heat map, which is the layout distribution of our gathered data set, using the suggested shadow guidance. The suggested system is data-driven, allowing users to analyze the design data naturally. The layout may then be edited by the user to finalize the layout design. We validated the suggested interface in our user study by comparing it with common design interfaces. The findings show that the suggested interface may achieve high retrieval accuracy while simultaneously offering a pleasant user experience.
\end{abstract}


\section{INTRODUCTION}
\label{sec:intro}  

Online courses and academic conferences are becoming popular and prevalent nowadays. In such online circumstances, the teachers and presenters may be required to create well-designed slides to give a lecture or make a presentation. The quality of the slides can affect the audience's comprehension \cite{garner2013design}. However, creating aesthetically pleasing presentations is time-consuming for inexperienced users. Although conventional software such as Microsoft PowerPoint is quite sophisticated, it may be more difficult for novel users to design well-designed slides than professional designers. Although PowerPoint includes a toolkit to automatically produce layout recommendations for users to assist them in designing slides, it has certain limitations including the inability to give a uniform style and make suggestions when there is too many materials on one single slide.

 \begin{figure} [t]
   \begin{center}
   \begin{tabular}{c}
   \includegraphics[width=0.85\textwidth]{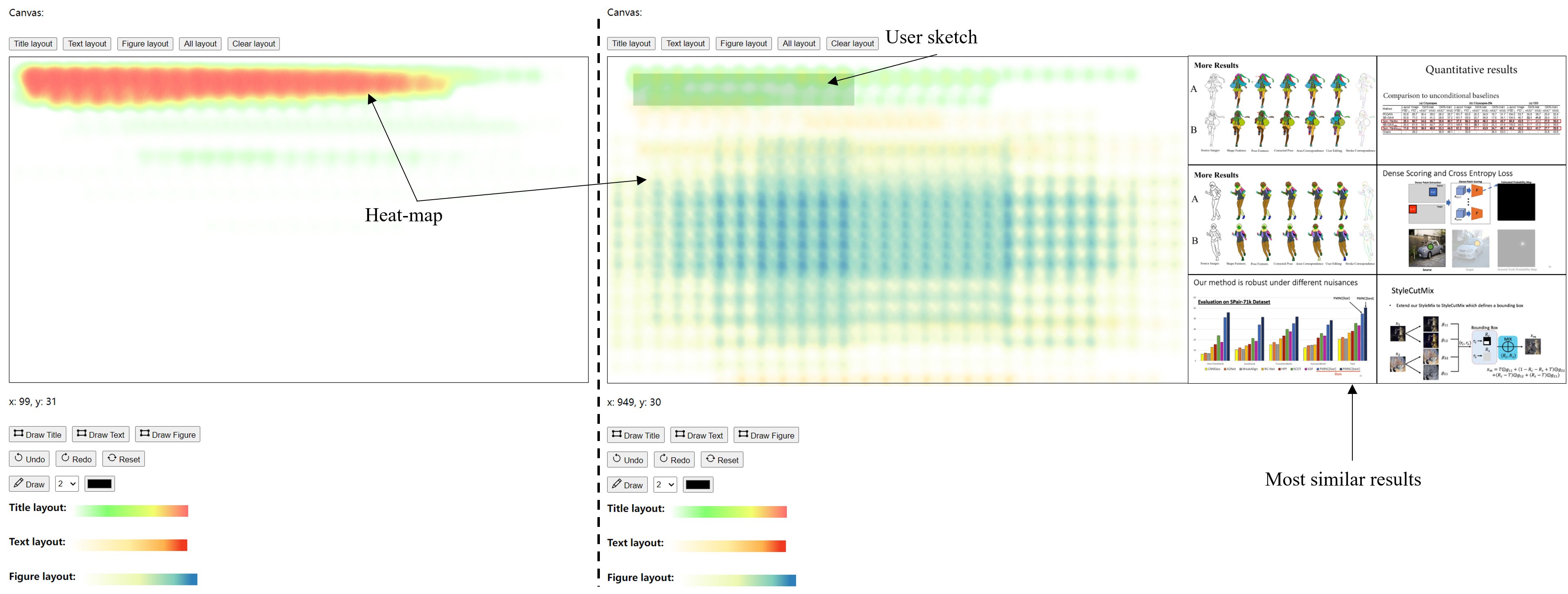}
    \end{tabular}
   \end{center}
   \caption[c] 
   {\label{fig:UI}  
The proposed interactive layout drawing interface. The left part is the original drawing canvas, the right part is the retrieval page after user draw the layout.}
 \end{figure} 
 
In this work, we propose an interactive drawing interface for ordinary users by retrieving the slide layout, as demonstrated in Figure \ref{fig:UI}. Because most academic presentations from online resource sites are in video format, it is difficult for users to acquire references to produce aesthetically appealing slides using the keyword-based retrieval method. Our proposed method extracts slides from the video input using a slide detection technique and determines the layout of the slides using a state-of-the-art image analysis approach. Then, when users retrieve slides, we use CNN (Convolutional Neural Network) to extract the features of the slides. Our system's interface is divided into three sections: the canvas section for user editing, the heatmap canvas section for shadow guidance, and the retrieval result section for displaying related slides with user editing. The design canvas can make it easy for users to search and find the reference slides. The heatmap canvas can motivate and excite unskilled users to consider how they might create the slide layout. The retrieval result page can provide guidance to users by selecting the desired layouts.

\section{RELATED WORK}
A drawing interface can use a free-hand drawing that is abstract and lacks visual details. The sketch can let users express their intention clearly. In such cases, some researches \cite{hashimoto2005retrieving} analyze layout sketches to retrieve the web page to give user references when designing, and other researches \cite{portenier2018faceshop,xiao2021sketchhairsalon} analyze the sketch stroke to edit the image. Previous studies \cite{huang2022dualface,lee2011shadowdraw} have also used user sketches to provide shadow guidance to enhance user design abilities, which can be used in motion retrieval~\cite{peng21} and calligraphy~\cite{he20}. The design interfaces have been proposed for cartoon image generation~\cite{luo21} and facial images~\cite{portenier2018faceshop}.  In this work, we aim to provide an interactive design interface to assist users in creating the layout of their slides.

VINS \cite{bunian2021vins} adopted the user interface layout as input to retrieve some designs of mobile interfaces to give user references when designing. Other research uses layout sketches to retrieve some example web pages that have similar layout designs \cite{hashimoto2005retrieving}. However, all of this research works only for web design. In addition, the other approaches used the neural network \cite{lee2020neural} and the transformer network \cite{arroyo2021variational} to generate the layout. In contrast to these approaches, our work focuses on supporting users to design rather than generating the layout automatically.

Deep learning based approaches were used extensively to analyze \cite{wu2019detectron2,zhong2019publaynet,xu2020layoutlm} and design layouts~\cite{lee2020neural,arroyo2021variational} . PubLayNet \cite{zhong2019publaynet}  contributed a huge dataset for document layout analysis. In this work, we utilize the deep learning method \cite{wu2019detectron2} to train a model for slides layout detection and analysis, and we apply the neural network \cite{simonyan2014very} to extract the feature of the slides layout.

\section{SYSTEM Overview}
In this work, we present an interactive layout retrieval interface and guide users in slide design.

\subsection{System Framework}
\begin{figure} [t]
   \begin{center}
   \begin{tabular}{c}
   \includegraphics[width=0.85\textwidth]{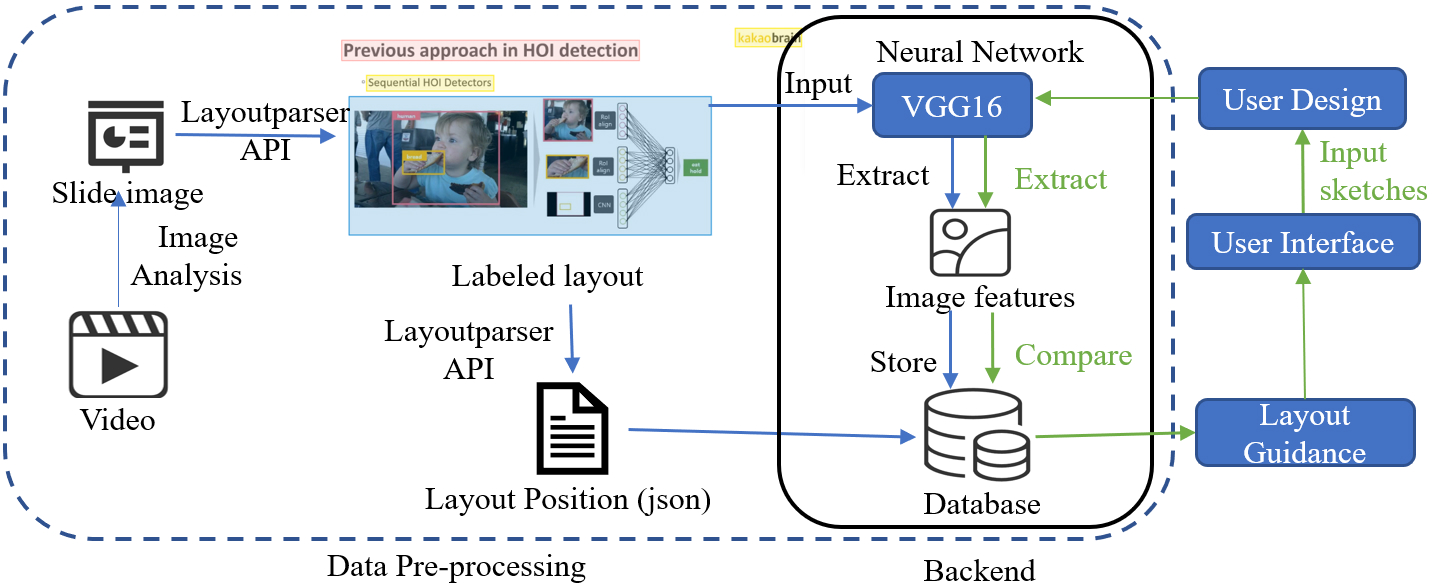}
    \end{tabular}
   \end{center}
   \caption[c] 
   {  \label{fig:FW}  
The framework of our system The left part is the offline computation process for slide extraction; the middle part is the dataset collection; and the right part is the user interface.}
 \end{figure}

The proposed system architecture consists of three parts: the offline computation process for slide extraction (pre-processing); the construction database considering slide layout features; and a layout drawing interface as shown in Figure \ref{fig:FW}. First, we extract slides from academic conference presentation videos and store them in the database. Then, we use the state-of-the-art layout analysis tool \cite{shen2021layoutparser} to analyze and label the layout of each slide in our database. After that, we use the Convolutional Neural Network (CNN) to extract the features of the labeled slide and store them in the database. The proposed interface consists of three sections: the drawing canvas section, the heat map canvas section, and the result section.

When the user draws a layout on the canvas, the client sends an HTTP request at the same time. The request is subsequently processed by the backend, which delivers image data to the client in the HTTP response. The data will then be rendered by the front-end, and the slide image will be displayed in the result area.

\subsection{Slides Extraction}
Slides extraction is crucial in this system because it is a fundamental stage in labeling layout, calculating distribution, and extracting the features of the slide layout, as illustrated in the left side of Figure \ref{fig:FW}. We begin by developing an algorithm to compare the image hash of each frame in a video of an academic conference presentation. If the difference between two frames is more than the threshold, the slide has changed and will be extracted. Following extraction, we computed the distribution of all slides and built a heat map, which is utilized to provide users with reference and inspiration. Detectron 2 \cite{wu2019detectron2} is then used to train a model to analyze the slides. Following that, we utilize deep learning based layoutparser (a tool that includes a collection of layout data structures and constructed APIs intended for document picture analysis tasks: picking layout components in the document and displaying the discovered layouts) to automatically label each slide's layout and CNN to extract the feature. This tool uses the model developed by Detectron 2 to extract difficult document structures using only a few lines of code and cutting-edge deep learning models. After the pre-processing is finished, all slide features are saved in the database and utilized for comparison with the user input designs.

\begin{figure} [t]
 \begin{center}
   \begin{tabular}{c}
\includegraphics[width=0.9\textwidth]{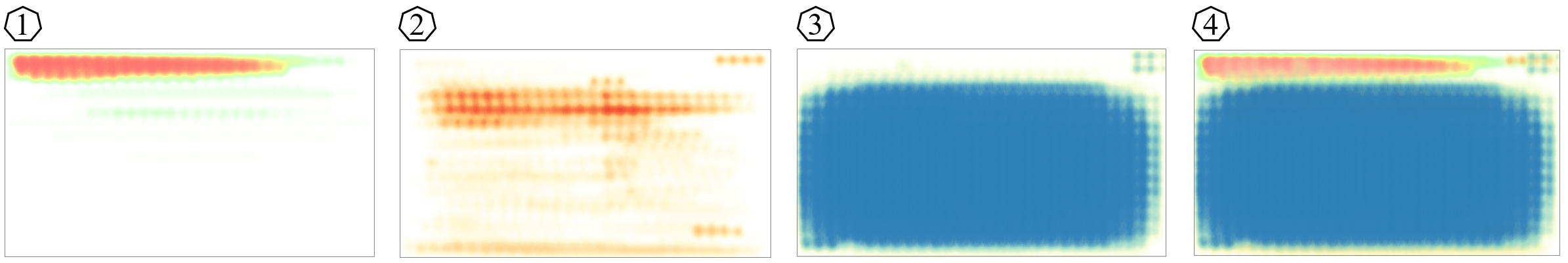}
\end{tabular}
   \end{center}
   \caption[c] 
{ \label{fig:HM} 
The shadow guidance of the proposed drawing interface. Part 1 is the heat-map of title layout, part 2 is the heat-map of text layout, part 3 is the heat-map of figure layout, and part 4 is the heat-map of all the layout.}
\end{figure}

\subsection{Drawing Interface}
The drawing interface can assist users in quickly and easily retrieving some related slides produced by themselves. It may also assist users in designing the slide layout and motivate them. When users want to acquire some ideas on how to create a slide layout, they may look at the heat-map (which indicates the distribution of each layout by different designers based on its color shades), which is displayed on the left side of Figure \ref{fig:UI}. Then the system will extract the features of the user input design and compare them with all the features stored in the database. The most similar slides will be rendered on the web page, as shown in the right part of Figure \ref{fig:UI}. When users edit their layouts, the reference result will change simultaneously.

As illustrated in Figure \ref{fig:HM}, the heat-map depicts the distribution of all slide layouts in the database. It was developed to provide users with some suggestions or inspiration for designing the layout. The heatmap was split into three sections: title, text, and figure. Most design scenarios can achieve this. The legend is located at the bottom of the heat-map canvas; the deeper the color, the larger the distribution. By pressing the various buttons, users can examine the distribution of each portion or the entire distribution, as illustrated in Figure \ref{fig:UI}. When users complete or alter the layout, the distribution of the heat-map changes concurrently to continually educate users on how to create the layout.

 \begin{figure} [ht]
   \begin{center}
   \begin{tabular}{c}
   \includegraphics[width=0.85\textwidth]{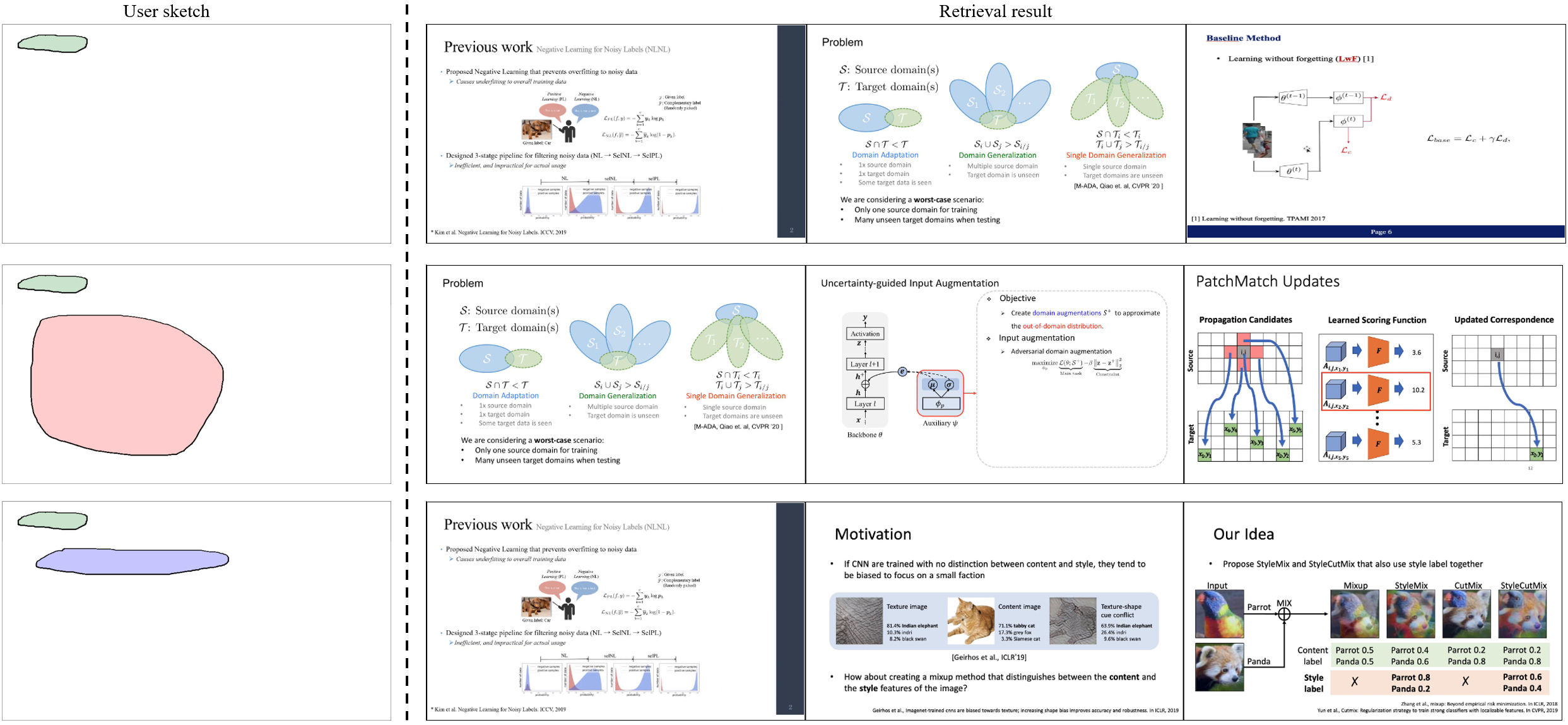}
    \end{tabular}
   \end{center}
   \caption[c] 
   {\label{fig:ldg} 
The retrieval results of layout design. The green box is the title layout, the red box is the figure layout, and the blue box is the text layout.}
 \end{figure} 
 
\section{User Study}
First, we conducted an experiment to compare the conventional retrieval interfaces with our interface. We invited 12 participants (college students around 20-years-old; 7 males and 5 females) into 2 groups, and another 5 participants (college students around 20-years-old, 5 males) to do the evaluation. After a brief introduction of how to use our interface, we let the first group use the traditional interface, PowerPoint (PPT), and the second group used our proposed interface. We first gave each group an academic paper (a poster in the field of computer science), and let each participant design three slides (only design the layout): introduction, method, and result. After they finish this task, we ask another 5 participants to evaluate their design (only evaluating the layout). The participant will assess three factors: the content in the slides is reasonably organized, the layout is designed aesthetically, and the slides are designed in a consistent style.
 
In the experiment about user experience, we conducted a user study to verify user experiences by asking participants to complete the questionnaire. The questionnaire uses a 7-Point Likert Scale (1 for strongly disagree and 7 for strongly agree). We asked the six participants who used our interface in the first experiment to complete the questionnaire.

\section{Results}
We discuss the implementation details, the result of the layout shadow guidance, the result of the user study, and user feedback in this section.

\subsection{Implementation Details}
In this work, the proposed interface was implemented in Python 3.8 as a website application on Windows 11. A personal computer with AMD Ryzen5 5600X (6-Core), 3.70 GHz, NVIDIA RTX3060 GPU, and 32GB RAM was used as the developing environment. In addition, we used Django 3.2 to establish the website and Vue 3.0 to render the website. We use the Detectron2 to train our model on our  dataset (443 slide images). Our prototype required 1.12s on average for retrieving the slides.

 \begin{figure} [ht]
   \begin{center}
   \begin{tabular}{c}
   \includegraphics[width=0.85\textwidth]{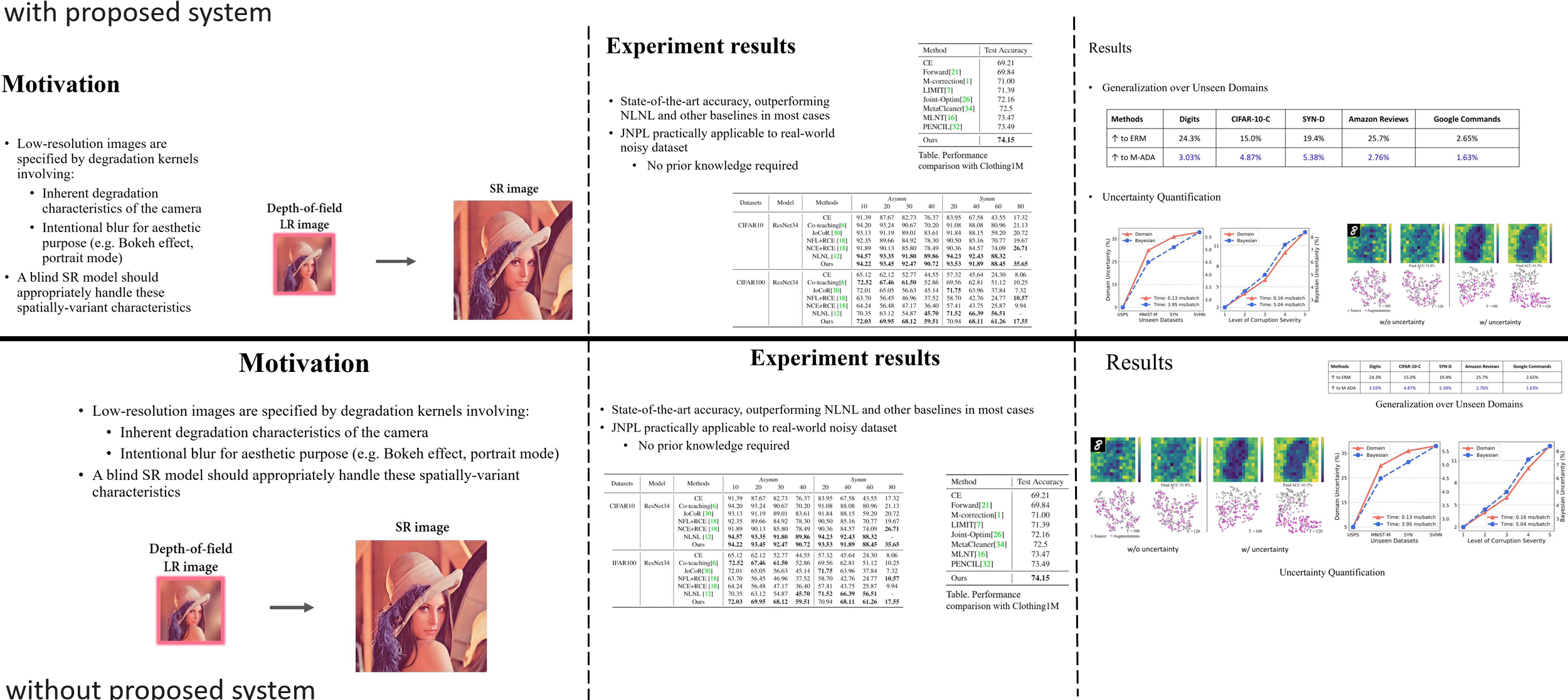}
    \end{tabular}
   \end{center}
   \caption[c] 
   {\label{fig:r} 
Examples of the designed results by participants during the user study.}
 \end{figure} 

\subsection{Layout Shadow Guidance}
Figure \ref{fig:ldg} shows the examples of retrieval results. Different input layouts can retrieve different slides. Even if users didn't complete the whole layout, the system will also retrieve the most similar slides for them according to their design. The upper part of Figure \ref{fig:ldg} shows that the system still works well with incomplete design input.

\subsection{User Study}
\begin{figure}[ht]
\begin{tabular}{c}
\begin{minipage}[t]{0.48\linewidth}
    \includegraphics[width = 1\linewidth]{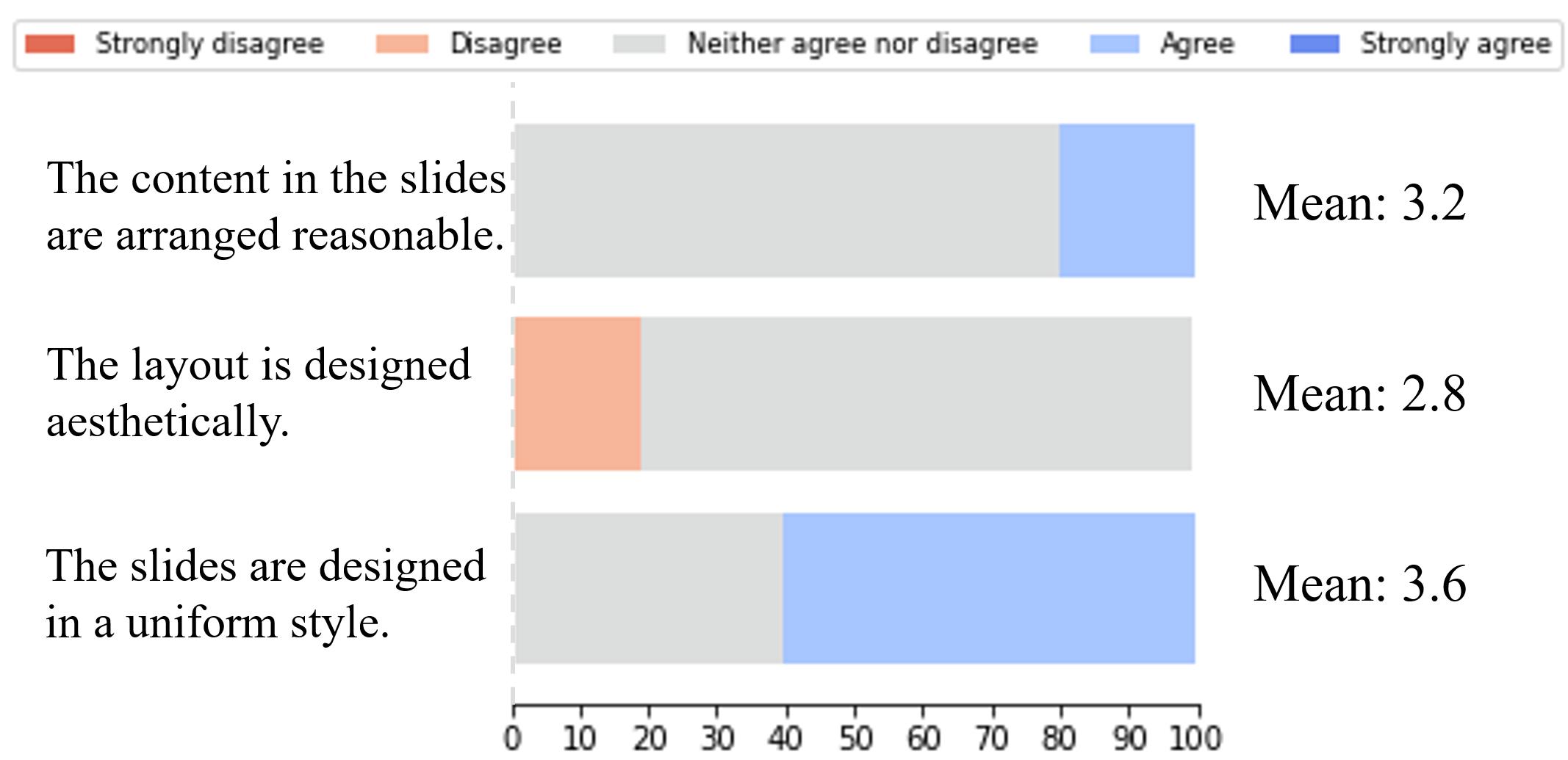}
    \caption[c] 
   {\label{fig:us1} 
The results of user study.}
\end{minipage}
\begin{minipage}[t]{0.48\linewidth}
    \includegraphics[width = 1\linewidth]{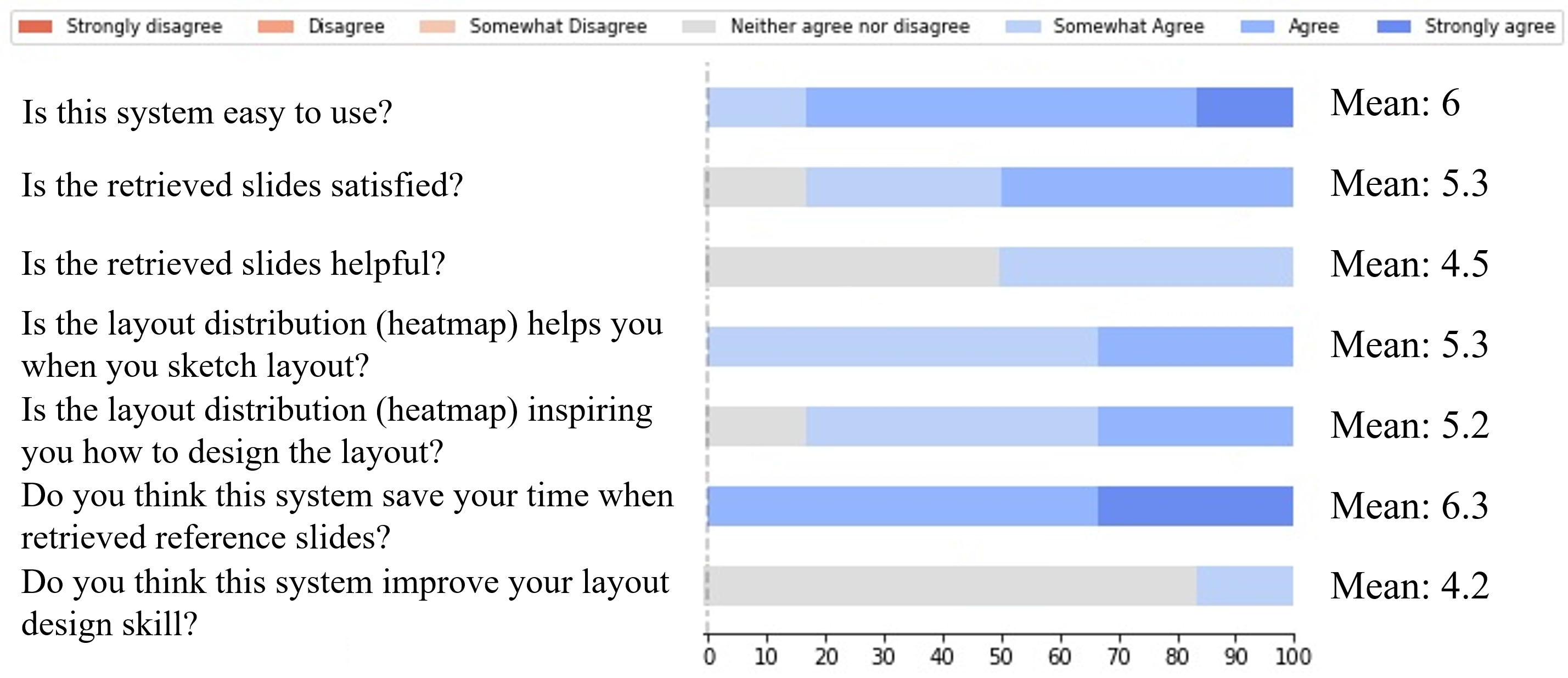}
    \caption[c] 
  { \label{fig:us2} 
The results of questionnaire.}
\end{minipage}
\end{tabular}
\end{figure}

The questionnaire for comparing experiment uses a 5-Point Likert Scale (1 for strongly disagreeing and 5 for strongly agreeing). The result is shown in the Figure \ref{fig:us1}. Our interface can provide users with satisfactory slides, whilst it can save their retrieval time. Furthermore, the design of a heat-map has been shown to inspire users on how to design layouts. Figure \ref{fig:r} shows some results designed by participants during the user study.

The result of user experience as shown in Figure \ref{fig:us2}. Although the proposed interface had difficulty to help users design slides with aesthetic appeal, we plan to extend our research to have the ability to provide references for the content of slides and add the well-designed slides to our database in the future. The user study verified that our interface can help users design more uniform styles than the conventional interface.

\section{Conclusion}
In this work, We proposed an layout drawing interface that can provide users with references and assist them in creating slide layouts. Using the proposed heatmap canvas can help and motivate users to make modifications to the layout. The users can design using the recommended references, which is thought to inspire the users to create unique layouts. The proposed interface can help novice users quickly find related resources and learn to construct the layout. According to the results of user study, the proposed interface can assist users create the slide in a consistent style and arrange the material more rationally than the traditional interface. It is also demonstrated that users prefer to access the reference through interactive modifications and that the reference might inspire them during the creation of the slide layout.

In the user study, some participants expressed dissatisfaction with the recovered presentations and claimed that the instruction falls short of their expectations. To solve this issue, we plan to expand the constructed database to include additional well-designed and high-quality presentations. We used 443 slides in current prototype implementation. It may be insufficient at times for users to obtain satisfactory outcomes. We only focused on layout design in this work, and we would like to broaden the current work to provide the content creations on the slide layouts.

\section*{Acknowledgements}
We thank all participants in our user study. This work was partially supported by the JAIST Research Grant and the JSPS KAKENHI Grant JP20K19845, Japan.

\bibliography{report} 
\bibliographystyle{spiebib} 

\end{document}